\documentclass[12pt]{article}
\usepackage{graphicx}
\usepackage{epstopdf}
\usepackage{epsfig}
\usepackage{amsmath}
\usepackage{multirow}
\usepackage{subfigure}
\usepackage{rotating}
\usepackage{array}
\usepackage{amsfonts}
\usepackage{afterpage}
\def\lsim{\mathrel{\raise.3ex\hbox{$<$\kern-.75em\lower 1ex\hbox{$\sim$}}}}
\def\gsim{\mathrel{\raise.3ex\hbox{$>$\kern-.75em\lower 1ex\hbox{$\sim$}}}}

\def\be{\begin{equation}}
\def\ee{\end{equation}}
\def\bea{\begin{eqnarray*}}
\def\eea{\end{eqnarray*}}

\begin{document}
\title{Texture and Cofactor Zeros of the Neutrino Mass Matrix}
\author{Jiajun Liao$^1$, D. Marfatia$^2$, and K. Whisnant$^1$\\
\\
\small\it $^1$Department of Physics and Astronomy, Iowa State University, Ames, IA 50011, USA\\
\small\it $^2$Department of Physics and Astronomy, University of Hawaii at Manoa, Honolulu, HI 96822, USA}
\date{}
\maketitle

\begin{abstract}

We study Majorana neutrino mass matrices 
that have two texture zeros, or two cofactor zeros, or one texture zero
and one cofactor zero. The two
texture/cofactor zero conditions give four constraints, which in conjunction with
 the five measured oscillation parameters completely determine the nine independent 
 real parameters of the neutrino mass matrix. We also study the implications that future 
measurements of neutrinoless double beta decay and the Dirac CP phase will have on these cases.

\end{abstract}

\newpage
\section{Introduction}
Neutrino phenomena are describable by the Majorana neutrino mass matrix, 
\begin{equation}
M=V^*\text{diag}(m_1, m_2, m_3)V^\dagger\,,
\label{eq:Mnu}
\end{equation}
where we work in the basis in which the charged lepton mass matrix is diagonal, the $m_i$ are real and nonnegative, $V=U\text{diag}(1, e^{i\phi_2/2}, e^{i\phi_3/2})$, and 
\begin{align}
U=\begin{bmatrix}
   c_{13}c_{12} & c_{13}s_{12} & s_{13}e^{-i\delta} \\
   -s_{12}c_{23}-c_{12}s_{23}s_{13}e^{i\delta} & c_{12}c_{23}-s_{12}s_{23}s_{13}e^{i\delta} & s_{23}c_{13} \\
   s_{12}s_{23}-c_{12}c_{23}s_{13}e^{i\delta} & -c_{12}s_{23}-s_{12}c_{23}s_{13}e^{i\delta} & c_{23}c_{13}
   \end{bmatrix}\,.
\label{eq:U}
\end{align}
Here $s_{ij}$ and $c_{ij}$ stand for the sine and cosine of the mixing angles $\theta_{ij}$.
We study the consequences of imposing two texture/cofactor zeros in the neutrino mass matrix. There are three classes of such ansatzes: two texture zeros (TT)~\cite{Frampton:2002yf, tt, Fritzsch:2011qv}, two cofactor zeros (CC)~\cite{Lavoura:2004tu, Lashin:2007dm}, and one texture zero and 
one cofactor zero (TC)~\cite{Dev:2010if}. Of the nine real parameters of $M$, five are fixed by measurements of the three mixing angles and two mass-squared differences; for a recent global three-neutrino fit see Ref.~\cite{Capozzi:2013csa}. The remaining four parameters, which we take to be the lightest mass, the Dirac phase, and the two Majorana phases, can then be determined from the four constraints that define the two texture/cofactor zeros.
Consequently, the rate for neutrinoless double beta decay ($0\nu\beta\beta$) which is given by the magnitude of the $\nu_e-\nu_e$
element of the neutrino mass matrix,
\begin{equation}
|M_{ee}| = |m_1c_{12}^2c_{13}^2 + m_2 e^{-i\phi_2}s_{12}^2c_{13}^2
+ m_3 e^{-i\phi_3}s_{13}^2e^{2i\delta}|\,,
\end{equation}
is also determined.

In Sections 2, 3 and 4, we use current experimental data to study the allowed
parameter space for two texture zeros, two cofactor zeros, and one texture and one cofactor
zero, respectively. We discuss and summarize our results in
Section 5.

\section{Two texture zeros}
The condition for a vanishing element $M_{\alpha\beta} = M_{\alpha\beta}^*= 0$ is
\begin{equation}
m_1 U_{\alpha 1} U_{\beta 1} + m_2 e^{i\phi_2}U_{\alpha 2} U_{\beta 2}
+ m_3 e^{i\phi_3} U_{\alpha 3} U_{\beta 3}=0\,.
\label{eq:texture}
\end{equation}
Since there are two such constraints that depend linearly on the masses, the masses are related by
\begin{equation}
\frac{m_1}{c_1} = \frac{m_2 e^{i\phi_2}}{c_2} = \frac{m_3e^{i\phi_3}}{c_3}\,,
\end{equation}
where $c_j$ are complex numbers that are quartic in the matrix elements of $U$. Then, with $\delta m^2 = m_2^2 - m_1^2$ and $\Delta m^2 = |m_3^2 - \frac{1}{2} (m_1^2 + m_2^2)|$, we get two equations that relate $m_1$ to the oscillation parameters and the Dirac phase $\delta$,
\begin{align}
m_1 &= \sqrt{\frac{\delta m^2 }{|c_2/c_1|^2 - 1}}\,,\label{eq:m1}
\\
m_1 &= \sqrt{\frac{\frac{1}{2}\delta m^2 \pm \Delta m^2 } {|c_3/c_1|^2 - 1}}\,,
\label{eq:m1texture}
\end{align}
where the plus and minus signs correspond to the normal hierarchy (NH) and the inverted hierarchy (IH), respectively. (For the NH the lightest mass is $m_1$, and for the IH the lightest mass is $m_3=\sqrt{m_1^2+\frac{1}{2}\delta m^2-\Delta m^2}$.) For a fixed set of oscillation parameters each of these two equations give $m_1$ as a function of $\delta$, and the intersections of the curves give the allowed values of $m_1$ and $\delta$. 
We use the data from the latest global fit of Ref.~\cite{Capozzi:2013csa} to find the $2\sigma$ allowed regions for the lightest mass and $\delta$ that satisfy Eqs.~(\ref{eq:m1}) and~(\ref{eq:m1texture}). Note that if we replace $\delta$ by $-\delta$, the two constraints from the Eqs.~(\ref{eq:m1}) and~(\ref{eq:m1texture}) will be the same since the magnitude of $c_i$ does not depend on the sign of $\delta$, but because the latest global fit has a preference for negative values of $\delta$~\cite{Capozzi:2013csa}, the allowed regions for $0\leq \delta \leq 180^\circ$ are a little larger than for $180^\circ\leq \delta \leq 360^\circ$.

For two texture zeros in the mass matrix, there are $\frac{6!}{2!4!} = 15$ different cases to consider. If two off-diagonal entries vanish, the mass matrices are block diagonal and have one neutrino decoupled from the others, which is inconsistent with the data. Therefore, we only need to consider 12 cases that can be divided into three categories:

\begin{table}
\centering
\begin{tabular}{|c|c|c|c|c|}\hline
Case & Structure & $c_1$ & $c_2$ & $c_3$ \\\hline
$X_1$ & $\left(\begin{array}{ccc} 0 & 0 &\times \\ 0&\times &\times \\ \times&\times&\times \end{array}\right)$ & $ U_{\tau 1}^*U_{e2}U_{e3}$ & $U_{\tau 2}^*U_{e3}U_{e1}$ & $U_{\tau 3}^*U_{e1}U_{e2}$ \\\hline
$X_2$ & $\left(\begin{array}{ccc} 0 & \times &0 \\ \times&\times &\times \\ 0&\times&\times \end{array}\right)$ & $ U_{\mu 1}^*U_{e2}U_{e3}$ & $U_{\mu 2}^*U_{e3}U_{e1}$ & $U_{\mu 3}^*U_{e1}U_{e2}$ \\\hline
$X_3$ & $\left(\begin{array}{ccc} \times & \times &\times \\ \times&\times &0 \\ \times &0 &0 \end{array}\right)$ & $ U_{e 1}^*U_{\tau 2}U_{\tau 3}$ & $U_{e 2}^*U_{\tau 3}U_{\tau 1}$ & $U_{e 3}^*U_{\tau 1}U_{\tau 2}$ \\\hline
$X_4$ & $\left(\begin{array}{ccc} \times & \times &\times \\ \times&0 &0 \\ \times &0 &\times \end{array}\right)$ & $ U_{e 1}^*U_{\mu 2}U_{\mu 3}$ & $U_{e 2}^*U_{\mu 3}U_{\mu 1}$ & $U_{e 3}^*U_{\mu 1}U_{\mu 2}$ \\\hline
$X_5$ & $\left(\begin{array}{ccc} \times & 0 &\times \\ 0&0 &\times \\ \times &\times &\times \end{array}\right)$ & $ U_{\tau 1}^*U_{\mu 2}U_{\mu 3}$ & $U_{\tau 2}^*U_{\mu 3}U_{\mu 1}$ & $U_{\tau 3}^*U_{\mu 1}U_{\mu 2}$ \\\hline
$X_6$ & $\left(\begin{array}{ccc} \times & \times &0 \\ \times&\times &\times \\ 0 &\times &0 \end{array}\right)$ & $ U_{\mu 1}^*U_{\tau 2}U_{\tau 3}$ & $U_{\mu 2}^*U_{\tau 3}U_{\tau 1}$ & $U_{\mu 3}^*U_{\tau 1}U_{\tau 2}$ \\\hline
\end{tabular}
\caption{The expressions for $c_1$, $c_2$ and $c_3$ for Class X. The symbol $\times$ denotes a nonzero matrix element.}
\label{tab:classX}
\end{table}

\begin{figure}
\centering
\includegraphics{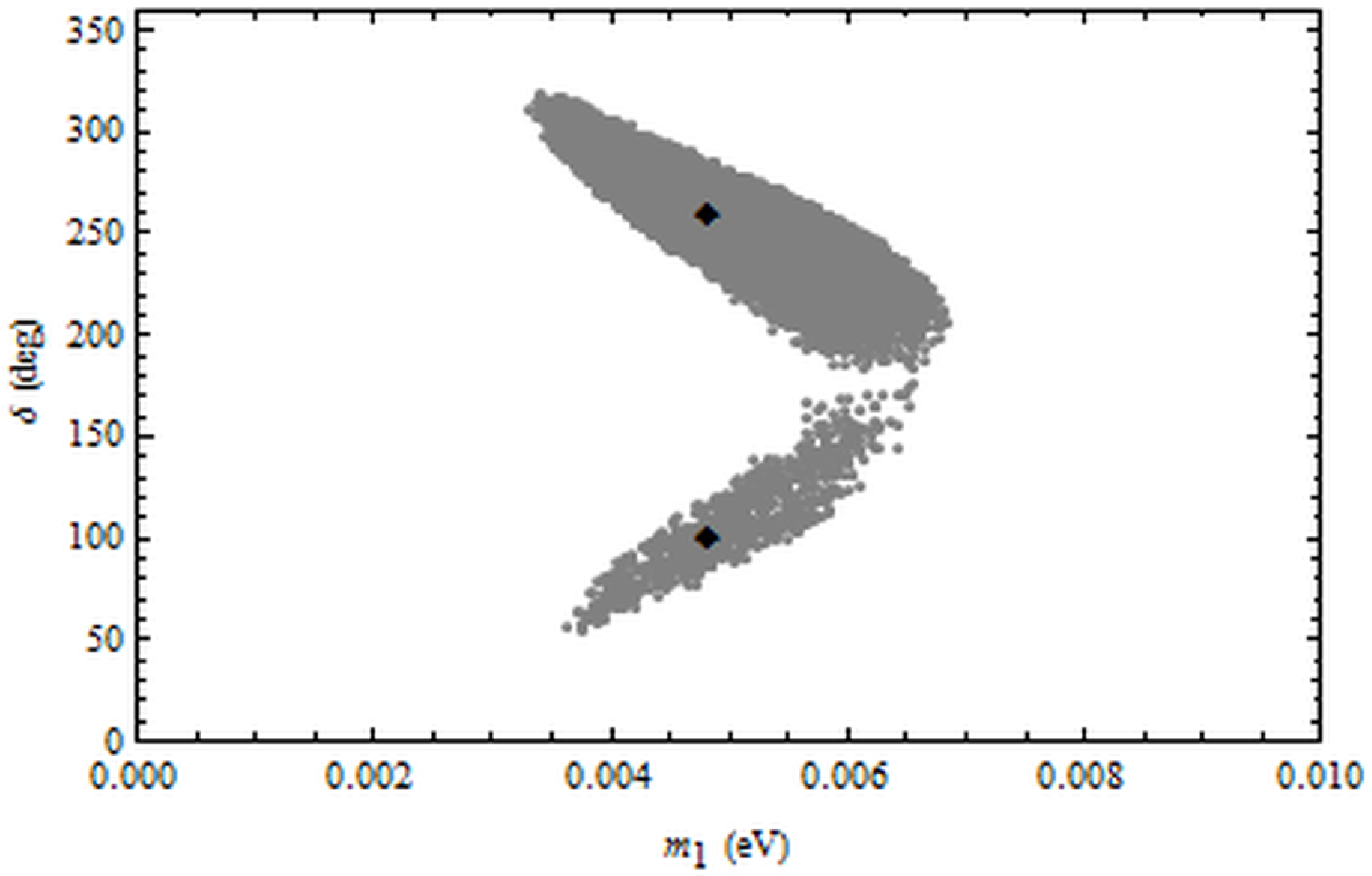}
\caption{The $2\sigma$ allowed regions in the $(m_1,\delta)$ plane for
the TT $X_1$ case and the normal hierarchy. The black diamonds indicate
$m_1$ and $\delta$ for  the best-fit values of the five oscillation
parameters.}
\label{fg:TT-X1-NH}
\end{figure}

\begin{figure}
\centering
\includegraphics{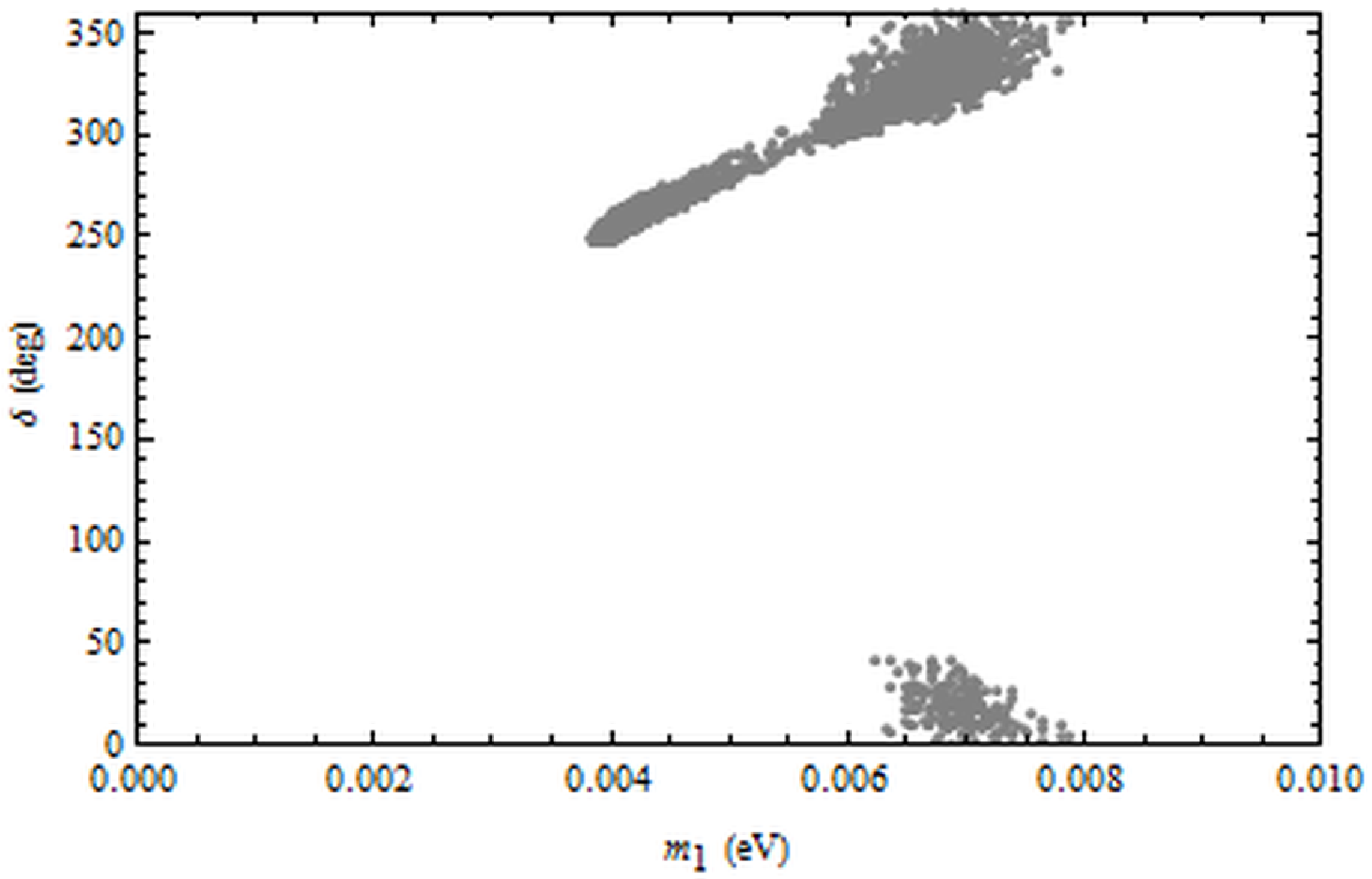}
\caption{Same as Fig.~\ref{fg:TT-X1-NH}, except for TT $X_2$ and
  the normal hierarchy. This case is not allowed for the best-fit
  oscillation parameters.}
\label{fg:TT-X2-NH}
\end{figure}

\begin{figure}
\centering
\includegraphics{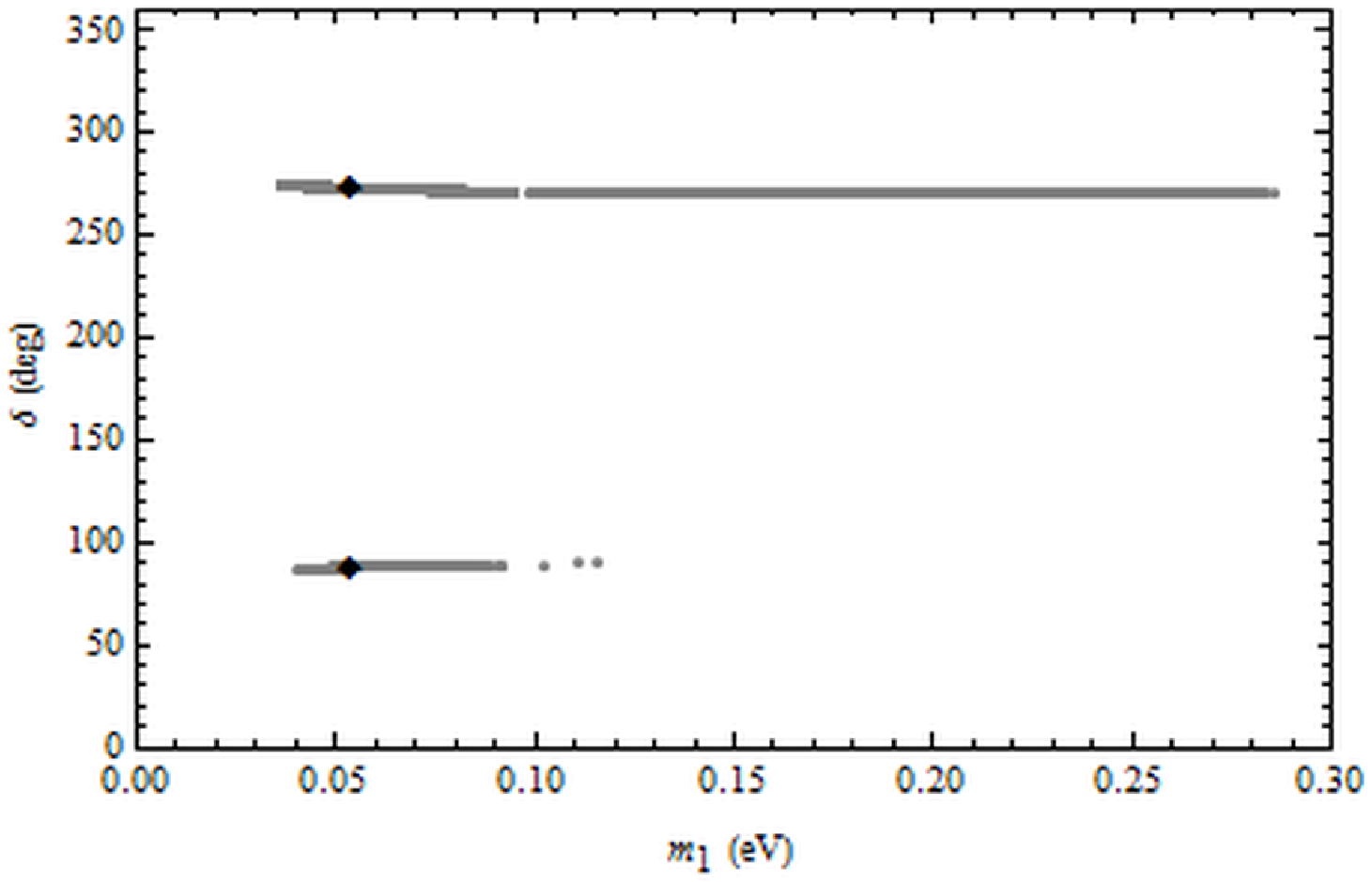}
\caption{Same as Fig.~\ref{fg:TT-X1-NH}, except for  TT $X_5$ and the normal hierarchy.}
\label{fg:TT-X5-NH}
\end{figure}

\begin{figure}
\centering
\includegraphics{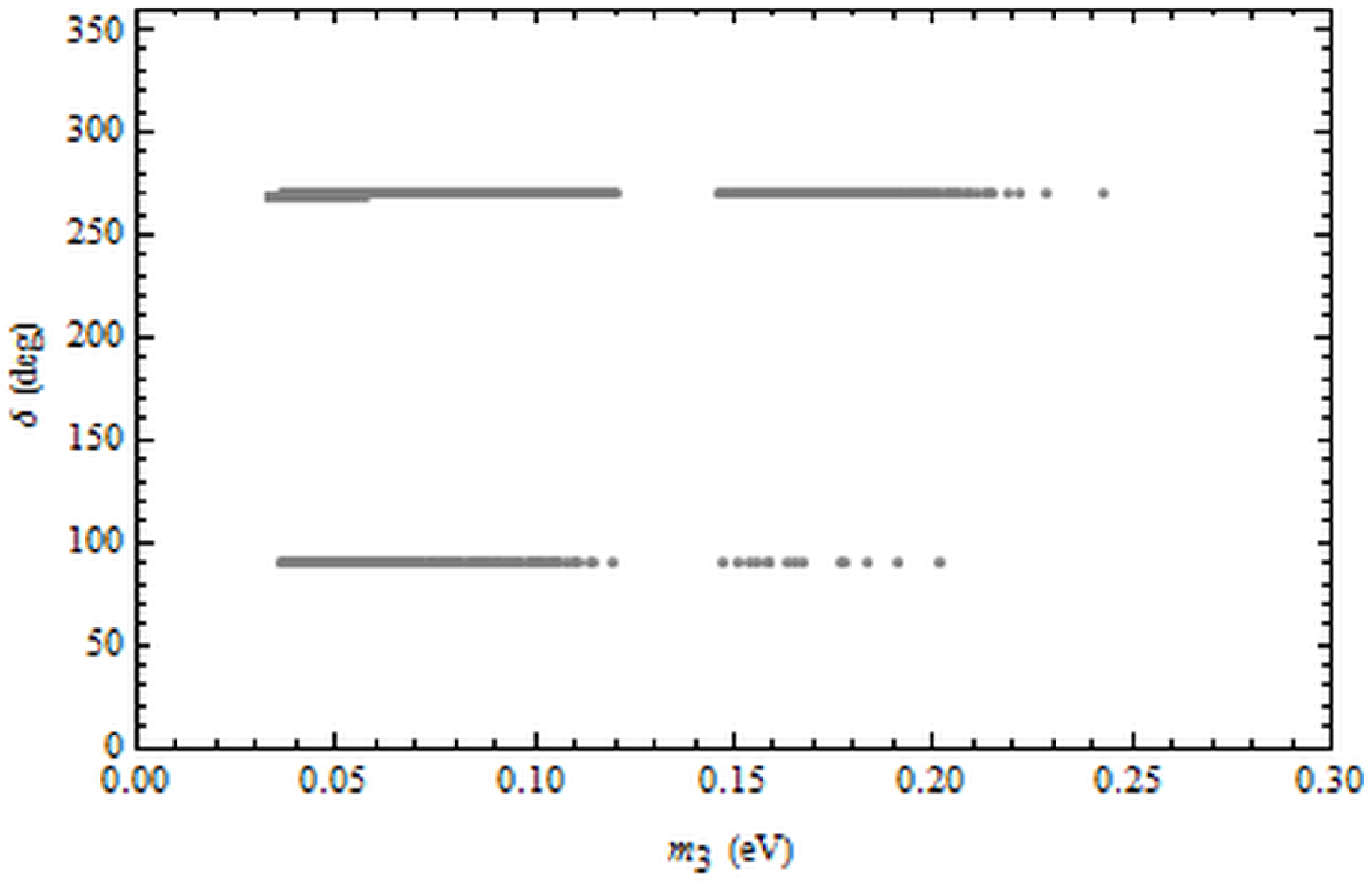}
\caption{Same as Fig.~\ref{fg:TT-X1-NH}, except for TT $X_5$ and
  the inverted hierarchy. This case is not allowed for the best-fit
  oscillation parameters.}
\label{fg:TT-X5-IH}
\end{figure}

\begin{enumerate}
\item \textbf{One zero on diagonal, off-diagonal zero sharing column
  and row.} The six possibilities of this type, $X_1$, $X_2$,
  $X_3$, $X_4$, $X_5$, and $X_6$, are displayed in Table~\ref{tab:classX}. Using the unitarity of $U$ and the
  fact that the cofactors of $U_{ij}$ are equal to $U_{ij}^*$, e.g.,
  $U_{e1} U_{\mu 2} - U_{e2} U_{\mu 1} = U_{\tau 3}^*$, we obtain
  the simplified expressions for $c_1$, $c_2$, and $c_3$ provided in Table~\ref{tab:classX}. 
From a numerical analysis, we find that at the $2\sigma$ level,
  only $X_1$, $X_2$ and $X_5$ are allowed for the normal hierarchy and
  $X_5$ and $X_6$ are allowed for the inverted hierarchy. The allowed regions for $X_1$ and $X_2$ for the
  normal hierarchy are shown in Figs.~\ref{fg:TT-X1-NH}
  and~\ref{fg:TT-X2-NH}. The allowed regions for $X_5$ for the normal
  and inverted hierarchy are shown in Figs.~\ref{fg:TT-X5-NH}
  and~\ref{fg:TT-X5-IH}, respectively. For the best-fit values of the
  measured oscillation parameters, $X_2$ NH and $X_5$ IH are not
  allowed, and the best-fit points for $X_1$ NH and $X_5$ NH are
  shown in Figs.~\ref{fg:TT-X1-NH} and~\ref{fg:TT-X5-NH}
  respectively. Both hierarchies for $X_5$ have
  nearly maximal CP violation, i.e., $\delta$ close to $90^\circ$ or
  $270^\circ$, and a lower bound on the lightest mass of about 30
  meV. For $X_5$ NH and $X_5$ IH, the upper bound on the
  lightest mass is about 290 meV and 250 meV, respectively. For comparison, the 95\%~C.L. limit from cosmology is
  $\sum m_i < 660$~meV~\cite{Reid:2009xm}. The allowed region for $X_6$ IH 
  is very similar to that for $X_5$ IH.

\item \textbf{One zero on diagonal, off-diagonal zero not sharing
  column and row.} The three possibilities of this type, $Y_1$,
  $Y_2$ and $Y_3$, and the corresponding $c_i$'s are displayed in
  Table~\ref{tab:classY}. At the $2\sigma$ level, $Y_1$ and $Y_2$ are
  allowed for the inverted hierarchy, and their allowed regions are
  very similar to that for $X_5$ IH; $Y_1$ is also allowed for the
  normal hierarchy and the allowed region is very similar to that for
  $X_5$ NH; $Y_3$ is excluded at $2\sigma$. All the allowed cases have
  nearly maximal CP violation, and a lower bound on the lightest mass
  of about 30 meV, similar to $X_5$ NH and $X_5$ IH; see Figs.~\ref{fg:TT-X5-NH} and~\ref{fg:TT-X5-IH}.

\item \textbf{Two zeros on diagonal.} The three possibilities of this type, $Z_1$, $Z_2$ and  $Z_3$, and the
corresponding $c_i$'s are listed in Table~\ref{tab:classZ}. The numerical results show that only $Z_1$ for the inverted hierarchy is allowed at the $2\sigma$ level, and the allowed regions are shown in Fig.~\ref{fg:TT-Z1-IH}. $Z_1$ for the normal hierarchy is excluded at $2\sigma$ for $m_1<0.3$ eV, which is consistent with the result of Ref.~\cite{Meloni:2014yea}.
\end{enumerate}

Although the allowed regions for the seven acceptable textures of Ref.~\cite{Frampton:2002yf} have been further restricted by the determination of $\theta_{13}$,
all seven textures remain allowed. Further restrictions on the Dirac CP phase $\delta$~\cite{Fritzsch:2011qv} can also be placed by the latest global fit~\cite{Capozzi:2013csa} for each case.

\begin{sidewaystable}
\centering
\begin{tabular}{|c|c|c|c|c|}\hline
Case & Structure & $c_1$ & $c_2$ & $c_3$ \\\hline
$Y_1$ & $\left(\begin{array}{ccc} \times & \times &0 \\ \times&0 &\times \\ 0 &\times &\times \end{array}\right)$ & $U_{e1}^*U_{e2}U_{\mu 3}-U_{\tau 1}^* U_{\mu 2}U_{\tau 3}$ & $U_{e 1}U_{e2}^*U_{\mu 3}-U_{\mu 1} U_{\tau 2}^*U_{\tau 3}$ & $U_{\mu 1}U_{e2}U_{e 3}^*-U_{\tau 1} U_{\mu 2}U_{\tau 3}^*$ \\\hline
$Y_2$ & $\left(\begin{array}{ccc} \times & 0 &\times \\ 0&\times &\times \\ \times &\times & 0 \end{array}\right)$ & $U_{\mu 1}^*U_{\tau 2}U_{\mu 3}-U_{e1}^* U_{e2}U_{\tau 3}$ & $U_{\tau 1}U_{\mu 2}^*U_{\mu 3}-U_{e1} U_{e2}^*U_{\tau 3}$ & $U_{\mu 1}U_{\tau 2}U_{\mu 3}^*-U_{\tau 1} U_{e2}U_{e 3}^*$ \\\hline
$Y_3$ & $\left(\begin{array}{ccc} 0 & \times &\times \\ \times&\times &0 \\ \times &0 &\times \end{array}\right)$ & $U_{\tau 1}^*U_{e 2}U_{\tau 3}-U_{\mu 1}^* U_{\mu 2}U_{e 3}$ & $U_{e 1}U_{\tau 2}^*U_{\tau 3}-U_{\mu 1}U_{\mu 2}^*U_{e 3}$ & $U_{\tau 1}U_{e 2}U_{\tau 3}^*-U_{e 1} U_{\mu 2}U_{\mu 3}^*$ \\\hline
\end{tabular}
\caption{The expressions for $c_1$, $c_2$ and $c_3$ for Class Y. The symbol $\times$ denotes a nonzero matrix element.}
\label{tab:classY}
\end{sidewaystable}

\begin{sidewaystable}
\centering
\begin{tabular}{|c|c|c|c|c|}\hline
Case & Structure & $c_1$ & $c_2$ & $c_3$ \\\hline
$Z_1$ & $\left(\begin{array}{ccc} \times & \times &\times \\ \times&0 &\times \\ \times &\times &0 \end{array}\right)$ & $(U_{\mu 2}U_{\tau 3}+U_{\mu 3}U_{\tau 2})U_{e1}^*$ & $(U_{\mu 3}U_{\tau 1}+U_{\mu 1}U_{\tau 3})U_{e2}^*$ & $(U_{\mu 1}U_{\tau 2}+U_{\mu 2}U_{\tau 1})U_{e3}^*$\\\hline
$Z_2$ & $\left(\begin{array}{ccc} 0 & \times &\times \\ \times& \times &\times \\ \times &\times &0 \end{array}\right)$ & $(U_{e 2}U_{\tau 3}+U_{e 3}U_{\tau 2})U_{\mu 1}^*$ & $(U_{e 3}U_{\tau 1}+U_{e 1}U_{\tau 3})U_{\mu 2}^*$ & $(U_{e 1}U_{\tau 2}+U_{e 2}U_{\tau 1})U_{\mu 3}^*$\\\hline
$Z_3$ & $\left(\begin{array}{ccc} 0 & \times &\times \\ \times&0 &\times \\ \times &\times &\times \end{array}\right)$ & $(U_{e 2}U_{\mu 3}+U_{e 3}U_{\mu 2})U_{\tau 1}^*$ & $(U_{e 3}U_{\mu 1}+U_{e 1}U_{\mu 3})U_{\tau 2}^*$ & $(U_{e 1}U_{\mu 2}+U_{e 2}U_{\mu 1})U_{\tau 3}^*$\\\hline
\end{tabular}
\caption{The expressions for $c_1$, $c_2$ and $c_3$ for Class Z. The symbol $\times$ denotes a nonzero matrix element.}
\label{tab:classZ}
\end{sidewaystable}

\section{Two cofactor zeros}
 In Ref.~\cite{Lashin:2007dm} it was shown that for matrices with two zero cofactors, the lightest mass can vanish only if $\theta_{13}=0$. Since $\theta_{13}$ is nonzero
 at  the $7.7\sigma$ level~\cite{An:2012eh}, we assume there are no vanishing neutrino masses and the mass matrix is invertible. Since $(M^{-1})_{\alpha\beta}=\frac{1}{\det M}C_{\beta\alpha}$ (where $C_{\alpha\beta}$ is the $(\alpha, \beta)$ cofactor of $M$), and the Majorana neutrino mass matrix is symmetric, $C_{\alpha\beta} = 0$ is equivalent to $(M^{-1})_{\alpha\beta}=0$. Because $M^{-1}=V\text{diag}(m_1^{-1}, m_2^{-1}, m_3^{-1})V^T$, the constraint is
\begin{equation}
m_1^{-1} U_{\alpha 1} U_{\beta 1} + m_2^{-1} e^{i\phi_2}U_{\alpha 2} U_{\beta 2}
+ m_3^{-1} e^{i\phi_3} U_{\alpha 3} U_{\beta 3}=0\,.
\label{eq:cofactor}
\end{equation}
The above equation is the same as Eq.~(\ref{eq:texture}), except
that the $m_i$'s are replaced by their inverses. Hence, we can follow
a procedure similar to that for the TT case to find the favored values of the lightest mass
and $\delta$. Since the cofactor
matrix is also diagonalized by the mixing matrix $V$, it 
cannot be block diagonal, and only 12 different patterns
need to be considered. It is possible to employ the notation for the
TT case if the locations of the two zeros are the same in the
cofactor matrix as in the mass matrix. Then all the $c_i$'s are identical, 
and the only difference from the TT case is that Eqs.~(\ref{eq:m1}) and~(\ref{eq:m1texture}) are replaced by
\begin{align}
m_1 &= \sqrt{\frac{\delta m^2 }{|c_1/c_2|^2 - 1}}\,,
\\
m_1 &= \sqrt{\frac{\frac{1}{2}\delta m^2 \pm \Delta m^2 } {|c_1/c_3|^2 - 1}}\,.
\end{align}
As for the TT case, there are three categories:

\begin{enumerate}
\item \textbf{One zero on diagonal, off-diagonal zero sharing column
  and row.} An interesting fact is that the two cofactor zero cases in
  this class yield the same allowed regions as for the two
  texture zero cases in the same
  class~\cite{Lavoura:2004tu,Lashin:2007dm}; the correspondence
  is listed in Table~\ref{tab:correspondence}. The reason for this is that
  the two cofactor zero conditions in this category imply either two texture zeros, or three cofactor
  zeros in a row or column. The latter possibility which gives a vanishing mass is
excluded since $\theta_{13} \neq 0$. 
From Table~\ref{tab:correspondence}, we readily find the cases that are allowed at $2\sigma$: $X_3$, $X_4$ and $X_6$  for the normal hierarchy, and $X_5$ and $X_6$ for
  the inverted hierarchy. 
The allowed regions in the $m_1$($m_3$)-$\delta$ plane are the same as those for the
 corresponding cases in the TT ansatz.

\begin{table}
\centering
\begin{tabular}{|c|c|c|c|c|c|c|}\hline
Two cofactor zeros&$X_1$&$X_2$&$X_3$&$X_4$&$X_5$&$X_6$\\\hline
Two texture zeros&$X_3$&$X_4$&$X_1$&$X_2$&$X_6$&$X_5$\\\hline
\end{tabular}
\caption{The correspondence between the two cofactor zero cases and two texture zero cases for Class X. }
\label{tab:correspondence}
\end{table}

\item \textbf{One zero on diagonal, off-diagonal zero not sharing
  column and row.} There are three possibilities of this type: $Y_1$,
  $Y_2$ and $Y_3$; see Table~\ref{tab:classY}. At the $2\sigma$ level, $Y_1$ and $Y_2$ are
  allowed for the inverted hierarchy, and their allowed regions are
  very similar to that for TT $X_5$ IH; $Y_2$ is also allowed for the
  normal hierarchy and the allowed region is very similar to that for
  TT $X_5$ NH; $Y_3$ is excluded at $2\sigma$. All the allowed cases
  have nearly maximal CP violation, and a lower bound on the lightest
  mass of about 30 meV, similar to TT $X_5$ NH and 
  and TT $X_5$ IH.

\item \textbf{Two zeros on diagonal.} There are three
  possibilities of this type: $Z_1$, $Z_2$ and $Z_3$; see Table~\ref{tab:classZ}. We find numerically
  that $Z_1$ is allowed at $2\sigma$ for the normal hierarchy only. The other cases are
  excluded at $2\sigma$. The allowed regions for $Z_1$ for the normal hierarchy
  are shown in Fig.~\ref{fg:CC-Z1-NH}.

\end{enumerate}

\begin{table}
\centering
\begin{tabular}{|c|c|}\hline
Case & Conditions\\\hline
$1A$&$M_{ee}=0,C_{ee}=0$\\\hline
$1B$&$M_{ee}=0,C_{e\mu}=0$\\\hline
$1C$&$M_{ee}=0,C_{e\tau}=0$\\\hline
$2A$&$M_{e\mu}=0,C_{ee}=0$\\\hline
$2D$&$M_{e\mu}=0,C_{\mu\mu}=0$\\\hline
$3A$&$M_{e\tau}=0,C_{ee}=0$\\\hline
$3F$&$M_{e\tau}=0,C_{\tau\tau}=0$\\\hline
$4B$&$M_{\mu\mu}=0,C_{e\mu}=0$\\\hline
$4D$&$M_{\mu\mu}=0,C_{\mu\mu}=0$\\\hline
$4E$&$M_{\mu\mu}=0,C_{\mu\tau}=0$\\\hline
$5D$&$M_{\mu\tau}=0,C_{\mu\mu}=0$\\\hline
$5F$&$M_{\mu\tau}=0,C_{\tau\tau}=0$\\\hline
$6C$&$M_{\tau\tau}=0,C_{e\tau}=0$\\\hline
$6E$&$M_{\tau\tau}=0,C_{\mu\tau}=0$\\\hline
$6F$&$M_{\tau\tau}=0,C_{\tau\tau}=0$\\\hline
\end{tabular}
\caption{The 15 cases with one texture zero and one cofactor zero that are not reducible to a TT case. }
\label{tab:tc}
\end{table}

\section{One texture zero and one cofactor zero}
There are 36 possibilities with one texture zero and one cofactor zero, of which 21 are equivalent to a TT case~\cite{Dev:2010if}. So we only need to study the remaining 15 cases listed in Table~\ref{tab:tc}. The two constraints $M_{\alpha\beta} = 0$ and $C_{\alpha'\beta'} = 0$ can be written as
\begin{equation}
m_1A_1+m_2e^{-i\phi_2}A_2+m_3e^{-i\phi_3}A_3=0\,,
\end{equation}
and
\begin{equation}
m_1^{-1}B_1+m_2^{-1}e^{i\phi_2}B_2+m_3^{-1}e^{i\phi_3}B_3=0\,,
\end{equation}
where $A_i=U_{\alpha i}^*U_{\beta i}^*$, and $B_i=U_{\alpha' i}U_{\beta' i}$ for $i=1,2,3$. Solving these two equations, we get
\begin{align}
\frac{m_3}{m_1}e^{-i\phi_3}&=\frac{1}{2A_3B_1}(A_2B_2-A_1B_1-A_3B_3\pm\sqrt{\Lambda})\,,\\
\frac{m_2}{m_1}e^{-i\phi_2}&=\frac{1}{2A_2B_1}(A_3B_3-A_1B_1-A_2B_2\mp\sqrt{\Lambda})\,,
\end{align}
where $\Lambda=A_1^2B_1^2+A_2^2B_2^2+A_3^2B_3^2-2(A_1A_2B_1B_2+A_1A_3B_1B_3+A_2A_3B_2B_3)$. Taking the absolute values of the above equations, we can find the two mass ratios, $\sigma=m_2/m_1$ and $\rho=m_3/m_1$. Then,
\begin{align}
m_1 &= \sqrt{\frac{\delta m^2 }{\sigma^2 - 1}},\nonumber
\\
m_1 &= \sqrt{\frac{\frac{1}{2}\delta m^2 \pm \Delta m^2 } {\rho^2 - 1}}\,,
\end{align}

\begin{figure}
\centering
\includegraphics{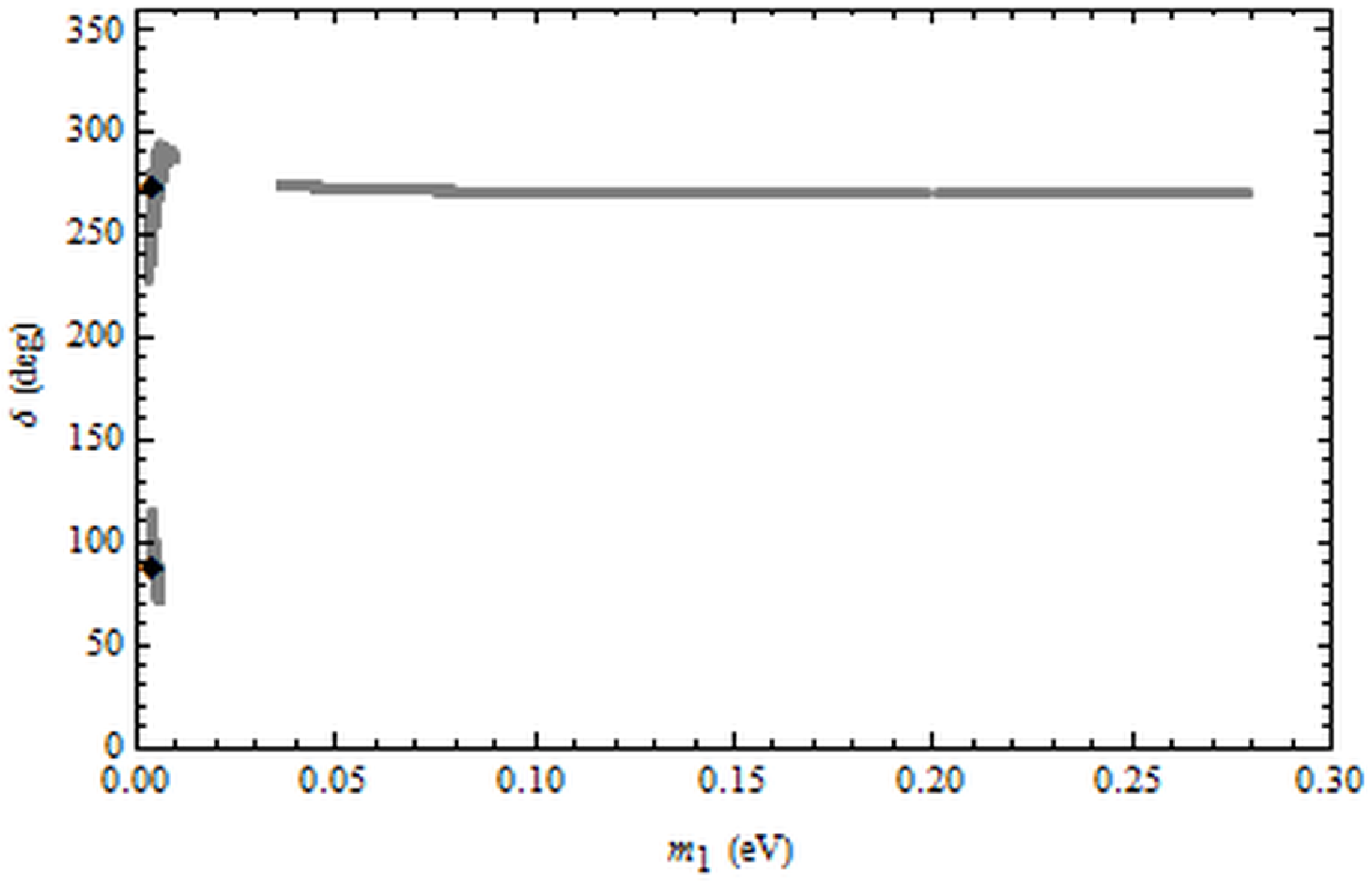}
\caption{Same as Fig.~\ref{fg:TT-X1-NH}, except for TC $2D$ and the normal hierarchy.}
\label{fg:TC-2D-NH}
\end{figure}

\begin{figure}
\centering
\includegraphics{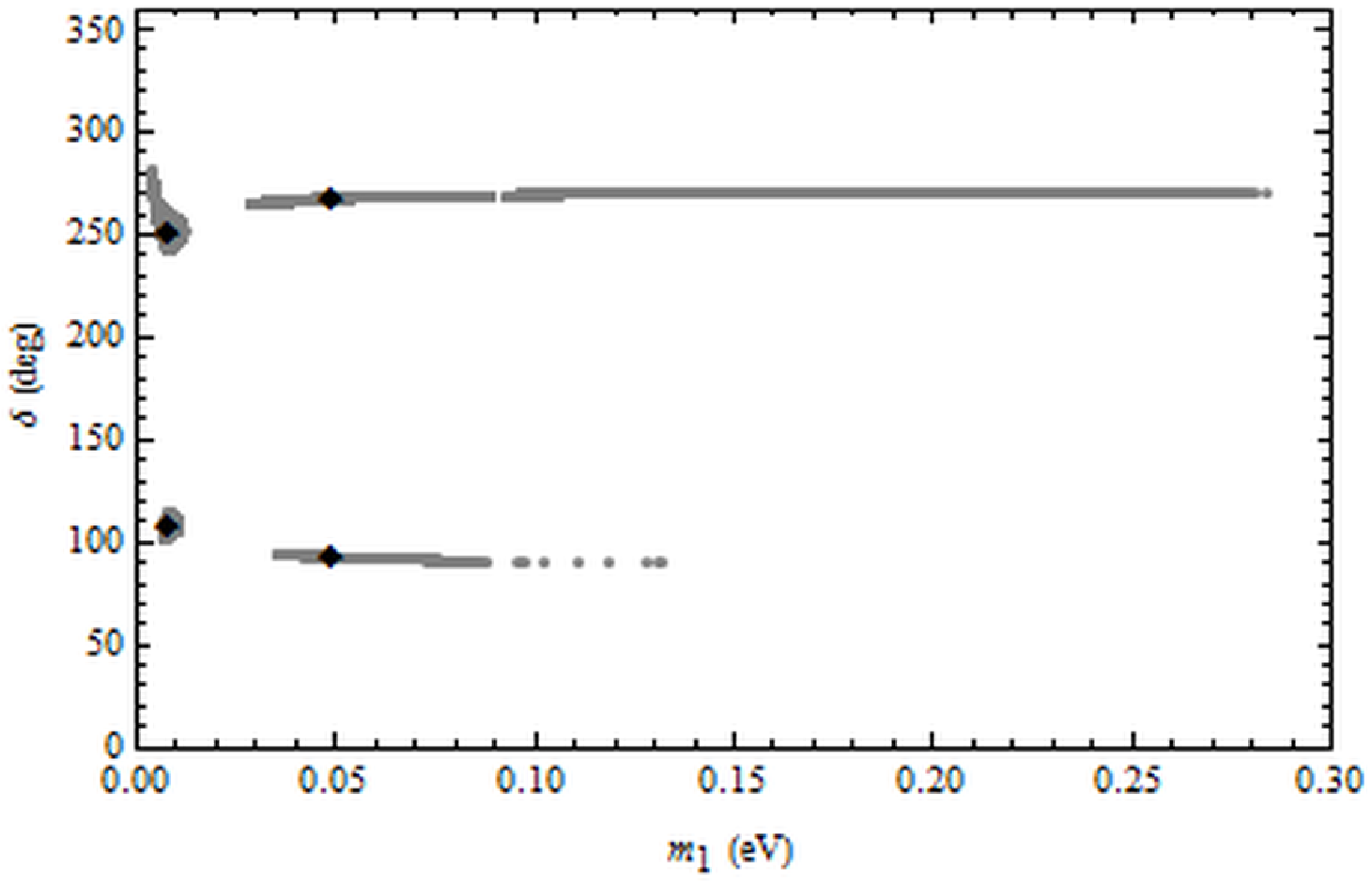}
\caption{Same as Fig.~\ref{fg:TT-X1-NH}, except for TC $3F$ and the normal hierarchy.}
\label{fg:TC-3F-NH}
\end{figure}

\begin{figure}
\centering
\includegraphics{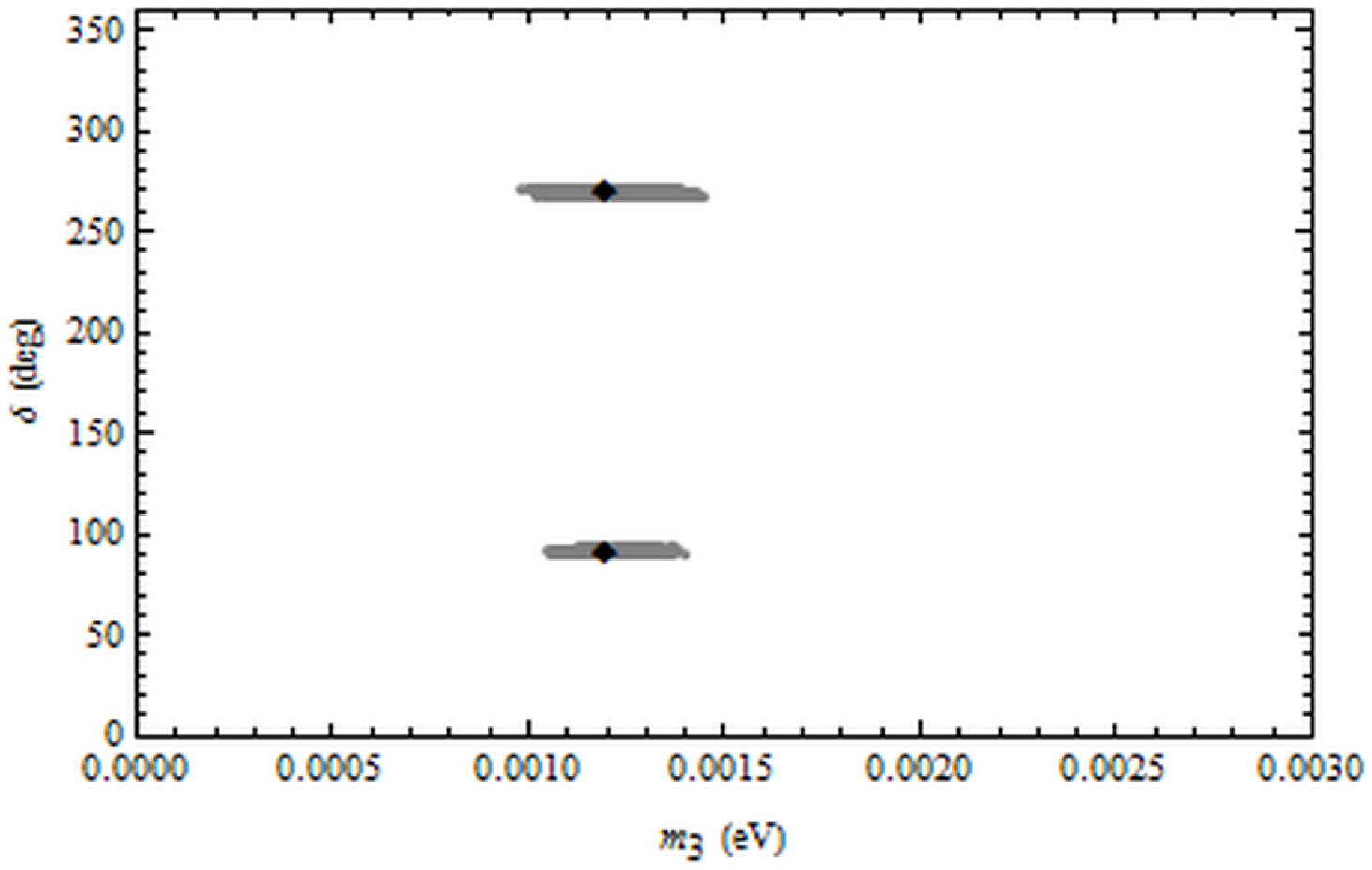}
\caption{Same as Fig.~\ref{fg:TT-X1-NH}, except for TC $2A$ and the inverted hierarchy.}
\label{fg:TC-2A-IH}
\end{figure}

\begin{figure}
\centering
\includegraphics{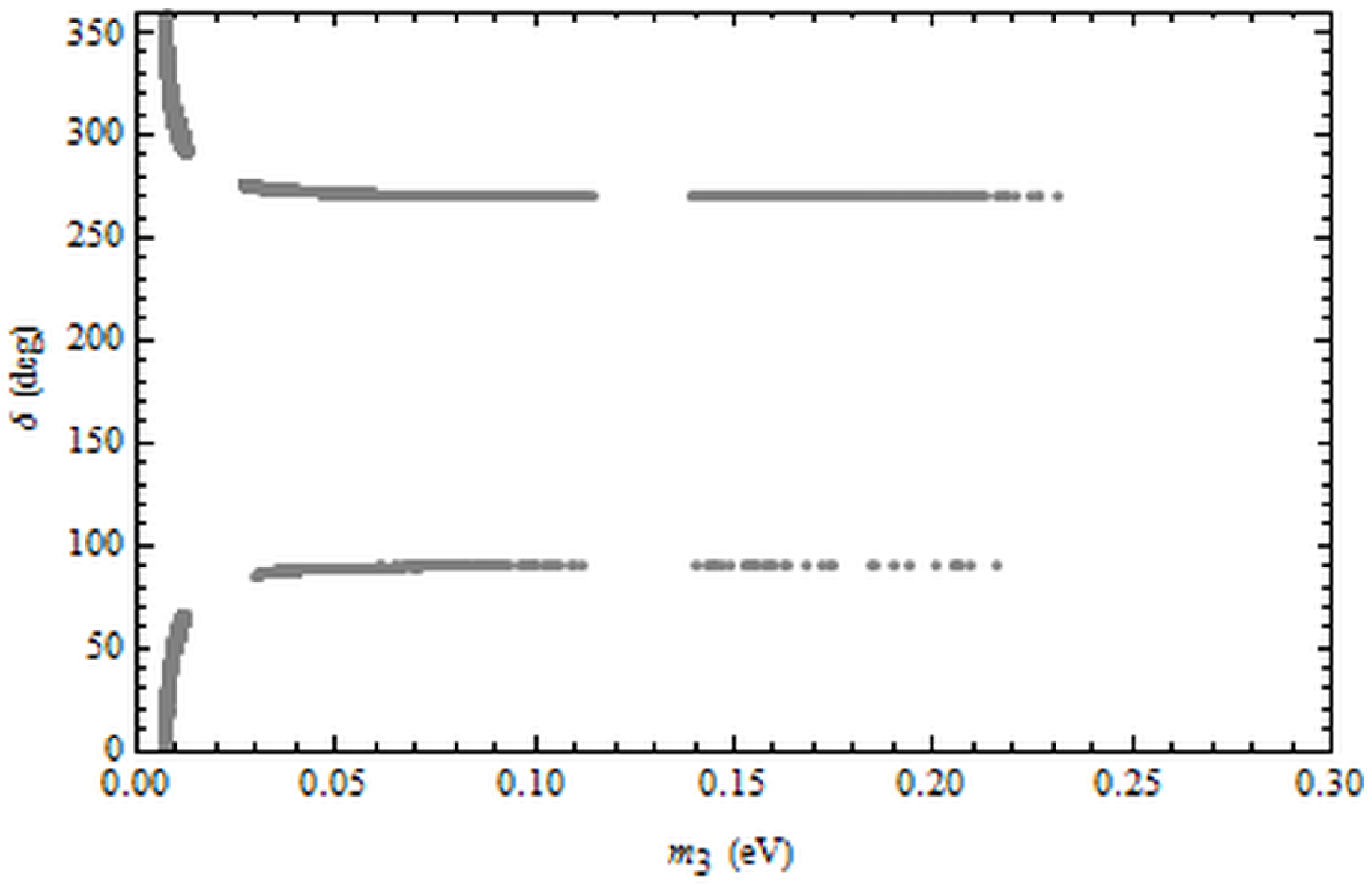}
\caption{Same as Fig.~\ref{fg:TT-X1-NH}, except for TC $4B$ and
  the inverted hierarchy. This case is not allowed for the best-fit
  oscillation parameters.}
\label{fg:TC-4B-IH}
\end{figure}

\begin{figure}
\centering
\includegraphics{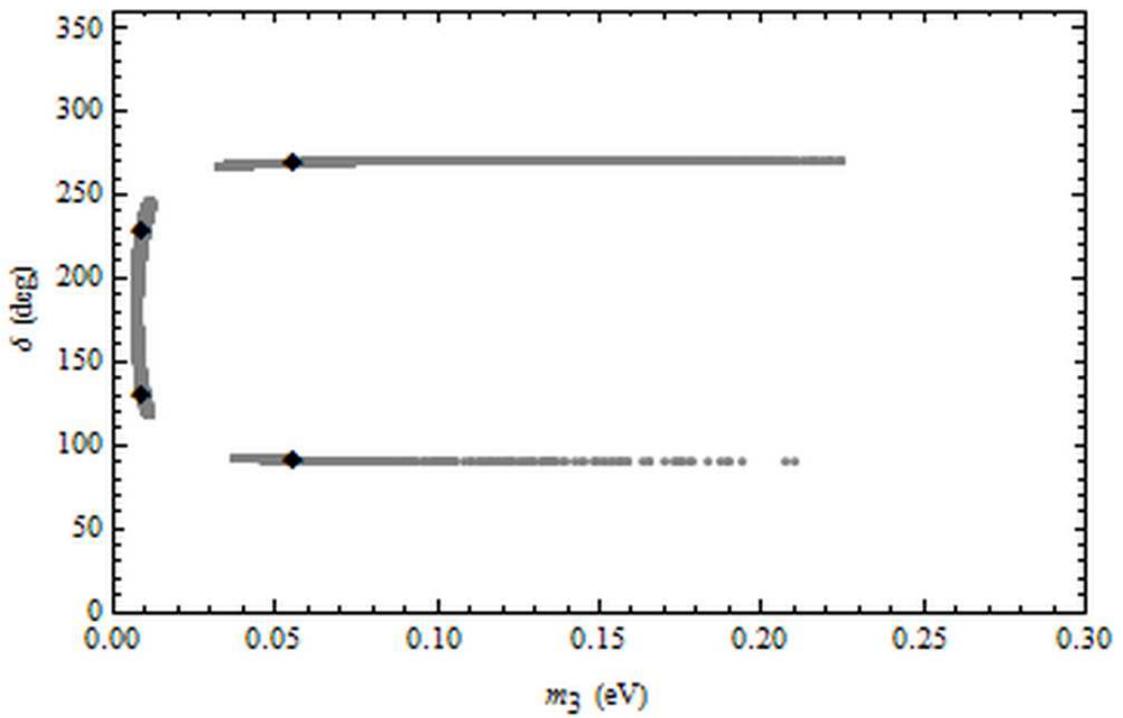}
\caption{Same as Fig.~\ref{fg:TT-X1-NH}, except for TC $6C$ and
  the inverted hierarchy. There are four best-fit points since there are
four solutions to the TC conditions.}
\label{fg:TC-6C-IH}
\end{figure}

A numerical study shows that at $2\sigma$, only $2D$, $3F$ and $4B$ are allowed 
for the normal hierarchy, and only $2A$, $2D$,
$3A$, $3F$, $4B$ and $6C$ are allowed for the
inverted hierarchy. The allowed regions for $2D$ and $3F$ for the
normal hierarchy are shown in Figs.~\ref{fg:TC-2D-NH} and~\ref{fg:TC-3F-NH},
and the allowed regions for $2A$, $4B$ and $6C$ for the inverted
hierarchy are shown in Figs.~\ref{fg:TC-2A-IH},~\ref{fg:TC-4B-IH}
and~\ref{fg:TC-6C-IH}, respectively. The allowed region for $3A$ IH is very similar to that for 
$2A$ IH.  The allowed region for
$4B$ NH is very similar to that for TT $X_5$ NH,
and the allowed regions for $2D$ IH and $3F$ IH 
are very similar to that for TT $X_5$ IH. They have nearly maximal CP
violation, and a lower bound on the lightest mass of about 30 meV. 
For $3F$ NH and $6C$
IH there are four best-fit points since there are four solutions to
the one texture zero and one cofactor zero conditions.

\section{Discussion}

There are 7 cases that are allowed at the $2\sigma$ level for the two
texture zero ansatz, 7 cases that are allowed for the two cofactor zero
ansatz, and 6 cases that are allowed for the one texture and one
cofactor zero ansatz. Seven cases allow both hierarchies, so there are a
total of 27 possible two-zero cases allowed at $2\sigma$. However,
there are many similarities among the allowed regions for these
cases. In Ref.~\cite{Liao:2013aka} we noted that any case with a
homogeneous relationship among elements of $M$
with one mass hierarchy yields predictions for the oscillation
parameters and phases similar to those given by a case with
the same homogeneous relationship among cofactors of $M$
with the opposite mass hierarchy. The only exceptions are when the
lightest mass is small, of order 20 meV or less, or when the allowed
ranges of the oscillation parameters differ significantly for the two
mass hierarchies. The latter situation occurs for $\theta_{23}$, which
is constrained at the $2\sigma$ level to be less than about $45.3^\circ$ for
the NH but can have larger values for the IH.

A texture or cofactor zero is the simplest homogeneous relationship;
therefore, CC cases can be dual to TT
cases (and, of course, vice versa), and some TC cases 
 can be dual to other such cases.\footnote{In
principle, the TC cases $1A$, $2D$, and $6F$ could be self-dual, which means
they would have similar allowed regions for the NH and IH, but these are
not allowed at $2\sigma$.} We can identify 8 cases where allowed
regions are similar due to the dual-case argument, 6 cases where a
case is allowed at $2\sigma$ but its dual case is not because the
lightest mass is small, 5 cases where an IH case is allowed but its
dual case is disfavored because $\theta_{23}$ must be larger than $45.3^\circ$.
A complete listing of dual case
relationships is given in Table~\ref{tab:dual}.

\begin{figure}
\centering
\includegraphics{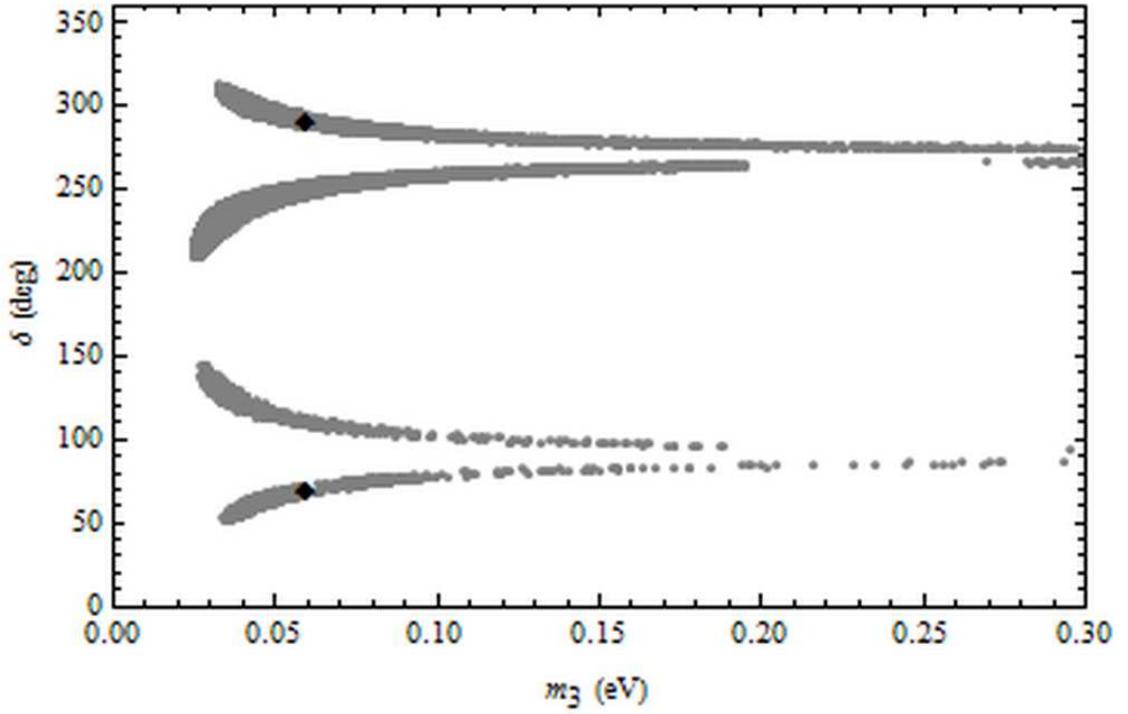}
\caption{Same as Fig.~\ref{fg:TT-X1-NH}, except for  TT $Z_1$ and the inverted hierarchy.}
\label{fg:TT-Z1-IH}
\end{figure}

\begin{figure}
\centering
\includegraphics{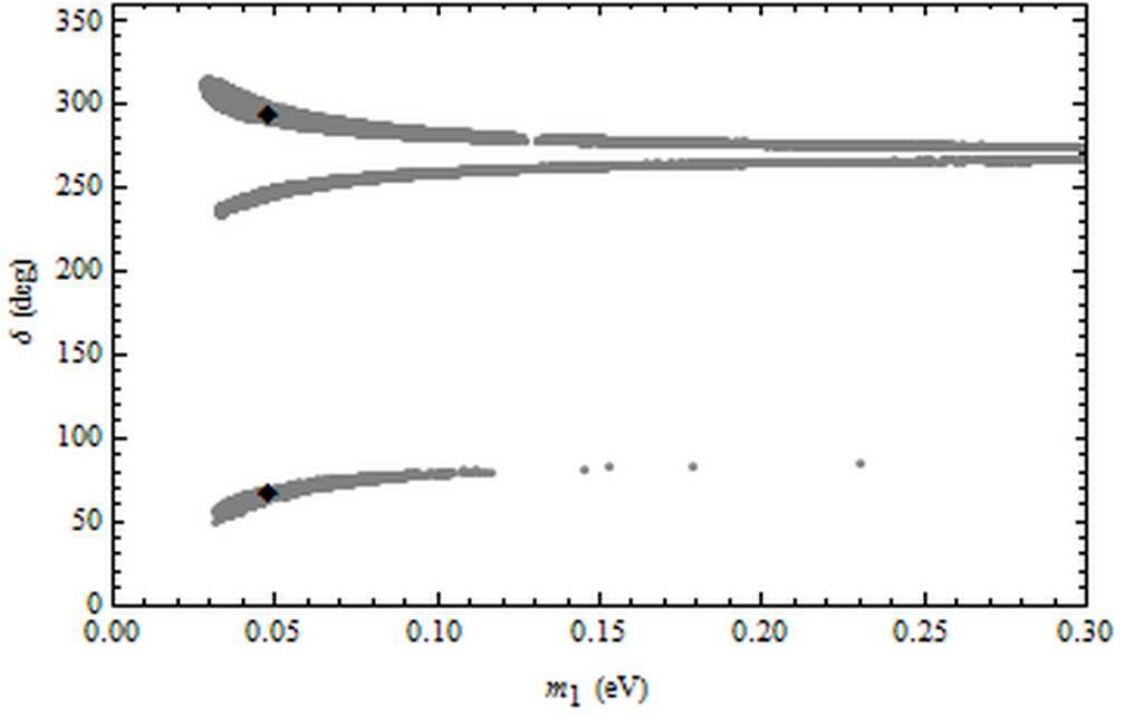}
\caption{Same as Fig.~\ref{fg:TT-X1-NH}, except for CC $Z_1$ and the normal hierarchy.}
\label{fg:CC-Z1-NH}
\end{figure}

We note that the allowed regions for the CC $Z_1$ NH case
(Fig.~\ref{fg:CC-Z1-NH}) are similar to the allowed
regions for its dual case, TT $Z_1$ IH (Fig.~\ref{fg:TT-Z1-IH}). The region for $90^\circ\le \delta \le 180^\circ$ in Fig.~\ref{fg:TT-Z1-IH} does not appear in
Fig.~\ref{fg:CC-Z1-NH} because $\theta_{23}$ has values that are larger than $45.3^\circ$ for the inverted hierarchy for $90^\circ\le\delta \le 180^\circ$, while such values of $\theta_{23}$ are disfavored for the normal hierarchy.

We can also use the effective Majorana mass for the neutrinoless double beta decay to differentiate two-zero cases. 
In Table~\ref{tab:mee}, we list the minimum and maximum values of
$|M_{ee}|$ at the $2\sigma$ level for each case. Note that
for TT $X_1$ and $X_2$, and CC $X_3$ and $X_4$, $|M_{ee}|$ is
identically zero, and therefore they are not listed in the table. We
also omit the cases of CC $X_5$ IH, $X_6$ NH and $X_6$ IH because they
give the same phenomenology as the corresponding cases in the TT
class, as given in Table~\ref{tab:correspondence}.

We find that there are four different types of cases phenomenologically:

\begin{enumerate}

\item \textbf{Cases that allow only a small value for the lightest
mass, less than 10~meV.} This includes 6 cases: TT $X_1$ NH, TT
$X_2$ NH, CC $X_3$ NH, CC $X_4$ NH, TC $2A$ IH and TC $3A$ IH (Figs.~\ref{fg:TT-X1-NH}, \ref{fg:TT-X2-NH} and~\ref{fg:TC-2A-IH}, respectively). The
value of $|M_{ee}|$ is either zero (for the TT cases) or close to 50~meV (for the TC cases).

\item \textbf{Cases that restrict $\delta$ to be very close to $90^\circ$
or $270^\circ$.} This group consists of 5 cases with NH (TT $X_5$,
CC $X_6$, TT $Y_1$, CC $Y_2$, and TC $4B$) and 10 cases with IH (TT
$X_5$, TT $X_6$, CC $X_5$, CC $X_6$, TT $Y_1$, CC $Y_1$, TT $Y_2$, CC
$Y_2$, TC $2D$, and TC $3F$), all of which have a minimum value for the
lightest mass of about 30~meV. In all NH cases of this type, the maximum
value for $m_1$ is about 290~meV and \mbox{$35\text{~meV}\lsim |M_{ee}| \lsim
290\text{~meV}$}; in all IH cases, the maximum value for $m_3$ is about
250~meV and $55\text{~meV}\lsim |M_{ee}| \lsim 250\text{~meV}$.
Therefore it will be very difficult to distinguish these cases from each
other.

\item \textbf{Cases in which maximal $CP$ violation is approached for
larger values of the lightest mass.} This group includes TT $Z_1$ IH and
CC $Z_1$ NH (Figs.~\ref{fg:TT-Z1-IH} and \ref{fg:CC-Z1-NH}). In these cases a wide range of $\delta$ is possible,
although maximal $CP$ violation is not allowed.

\item \textbf{Cases that are a mixture of types 1 and 2.} The TC
cases $2D$ NH, $3F$ NH, $4B$ IH, and $6C$ IH allow values of the lightest mass less
than 10~meV, and also have nearly maximal $CP$ violation when the
lightest mass is above about 30~meV (Figs.~\ref{fg:TC-2D-NH}, \ref{fg:TC-3F-NH}, \ref{fg:TC-4B-IH},
and \ref{fg:TC-6C-IH}).

\end{enumerate}

\begin{table}
\begin{center}
\begin{tabular}{|c|c|p{10cm}|}\hline
Case & Hierarchy & Dual case allowed?\\\hline
TT $X_1$ & NH & No, $m_1$ small\\\hline
TT $X_2$ & NH & No, $m_1$ small\\\hline
TT $X_5$ & NH & Yes, CC $X_5$ IH\\\hline
TT $X_5$ & IH & Maybe, CC $X_5$ NH, if the $\theta_{23}$ restriction were absent\\\hline
TT $X_6$ & IH & Yes, CC $X_6$ NH\\\hline
TT $Y_1$ & NH & Yes, CC $Y_1$ IH\\\hline
TT $Y_1$ & IH & Maybe, CC $Y_1$ NH, if the $\theta_{23}$ restriction were absent\\\hline
TT $Y_2$ & IH & Yes, CC $Y_2$ NH\\\hline
TT $Z_1$ & IH & Yes, CC $Z_1$ NH (for $\delta\in [0,90^\circ]\cup [180^\circ,360^\circ]$)\\\hline
CC $X_3$ & NH & No, $m_1$ small\\\hline
CC $X_4$ & NH & No, $m_1$ small\\\hline
CC $X_6$ & IH & Maybe, TT $X_6$ NH, if the $\theta_{23}$ restriction were absent\\\hline
CC $Y_2$ & IH & Maybe, TT $Y_2$ NH, if the $\theta_{23}$ restriction were absent\\\hline
TC $2A$ & IH & No, $m_3$ small\\\hline
TC $2D$ & NH & Yes, TC $4B$ IH (except for low $m_1$ and $m_3$ and if the $\theta_{23}$ restriction were absent)\\\hline
TC $2D$ & IH & Yes, TC $4B$ NH\\\hline
TC $3A$ & IH & No, $m_3$ small\\\hline
TC $3F$ & NH & Yes, TC $6C$ IH (except for low $m_1$ and $m_3$)\\\hline
TC $3F$ & IH & Maybe, TC $6C$ NH, if the $\theta_{23}$ restriction were absent\\\hline
\end{tabular}
\end{center}
\caption{A listing of which allowed cases have dual cases that are also
allowed, and which do not. The ``Maybe" designation is for situations in which the dual case has a NH and $\theta_{23}>45.3^\circ$; the global analysis of Ref.~\cite{Capozzi:2013csa} suggests that for a NH, $\theta_{23}<45.3^\circ$ at 2$\sigma$.  ``Maybe" indicates that the exclusion of the dual case on this basis is not robust.}
\label{tab:dual}
\end{table}

\begin{table}
\begin{center}
\begin{tabular}{|c|c|c|c|}\hline
Case&Hierarchy&Minimum&Maximum\\\hline
TT $X_5$&NH&37&286\\\hline
TT $X_5$&IH&58&247\\\hline
TT $X_6$&IH&62&215\\\hline
TT $Y_1$&NH&34&276\\\hline
TT $Y_1$&IH&57&230\\\hline
TT $Y_2$&IH&60&226\\\hline
TT $Z_1$&IH&24&$>$1000\\\hline
CC $Y_1$&IH&59&231\\\hline
CC $Y_2$&NH&34&275\\\hline
CC $Y_2$&IH&56&227\\\hline
CC $Z_1$&NH&15&$>$1000\\\hline
TC $2A$&IH&45&49\\\hline
TC $2D$&NH&3&279\\\hline
TC $2D$&IH&60&227\\\hline
TC $3A$&IH&45&49\\\hline
TC $3F$&NH&3&281\\\hline
TC $3F$&IH&57&217\\\hline
TC $4B$&NH&35&281\\\hline
TC $4B$&IH&16&232\\\hline
TC $6C$&IH&15&229\\\hline
\end{tabular}
\end{center}
\caption{The minimum and maximum values of $|M_{ee}|$ (in meV) 
at  $2\sigma$. CC $X_5$ and CC $X_6$ 
are not shown since they are equivalent to TT $X_6$ and TT $X_5$,
respectively.  $|M_{ee}|$ is identically zero for TT $X_1$, TT $X_2$, CC $X_3$ and CC $X_4$.}
\label{tab:mee}
\end{table}

Due to the large number of cases and their overlapping predictions, it is currently not possible to uniquely determine any given case. The latest experimental result from EXO-200~\cite{Auger:2012ar} sets an upper limit on the effective mass $|M_{ee}|$ of less than $140-380$ meV at 90\% C.L. However, with future sensitivities to $|M_{ee}|$ of about 20~meV~\cite{Rodejohann:2012xd}, and a precision
measurement of $\delta$ in future long baseline oscillation experiments, we might be able to distinguish between these cases. Here we run a test on the survivability of two-zero cases by applying an upper limit on $|M_{ee}|$ and assuming specific values for $\delta$ with the 3$\sigma$ resolution attainable with a 350~kt-yr exposure at the Long-Baseline Neutrino Experiment~\cite{Adams:2013qkq}. The results in Table~\ref{tab:survival} are qualitative without specific
confidence levels ascribable.
%
%
%
%
\begin{table}
\caption{The two-zero cases that survive (indicated by a tick mark) an upper limit on $|M_{ee}|$ and a measurement of $\delta$ (as in the second row)  with the 3$\sigma$ resolution attainable by the Long-Baseline Neutrino Experiment with 350~kt-yr of data~\cite{Adams:2013qkq}. The CC Class X is not shown since it is equivalent to the TT Class X.}
\begin{center}
\begin{tabular}{|c|c|c|c|c|c|c|c|c|c|c|c|c|}\hline
\multirow{2}{*}{Case}&\multicolumn{4}{c|}{$|M_{ee}|<20$~meV}&\multicolumn{4}{c|}{$|M_{ee}|<50$~meV}&\multicolumn{4}{c|}{$|M_{ee}|<100$~meV}\\\cline{2-13}
                      &0&$90^\circ$&$180^\circ$&$270^\circ$&0&$90^\circ$&$180^\circ$&$270^\circ$&0&$90^\circ$&$180^\circ$&$270^\circ$\\\hline
TT $X_1$ NH &$\times$&$\surd$&$\surd$&$\surd$ &$\times$&$\surd$&$\surd$&$\surd$ &$\times$&$\surd$&$\surd$&$\surd$\\\hline 
TT $X_2$ NH &$\surd$&$\surd$&$\times$&$\surd$ &$\surd$&$\surd$&$\times$&$\surd$ &$\surd$&$\surd$&$\times$&$\surd$\\\hline 
TT $X_5$ NH &$\times$&$\times$&$\times$&$\times$ &$\times$&$\surd$&$\times$&$\surd$ &$\times$&$\surd$&$\times$&$\surd$\\\hline 
TT $X_5$ IH &$\times$&$\times$&$\times$&$\times$ &$\times$&$\times$&$\times$&$\times$ &$\times$&$\surd$&$\times$&$\surd$\\\hline 
TT $X_6$ IH &$\times$&$\times$&$\times$&$\times$ &$\times$&$\times$&$\times$&$\times$ &$\times$&$\surd$&$\times$&$\surd$\\\hline 
TT $Y_1$ NH &$\times$&$\times$&$\times$&$\times$ &$\times$&$\surd$&$\times$&$\surd$ &$\times$&$\surd$&$\times$&$\surd$\\\hline
TT $Y_1$ IH &$\times$&$\times$&$\times$&$\times$ &$\times$&$\times$&$\times$&$\times$ &$\times$&$\surd$&$\times$&$\surd$\\\hline 
TT $Y_2$ IH &$\times$&$\times$&$\times$&$\times$ &$\times$&$\times$&$\times$&$\times$ &$\times$&$\surd$&$\times$&$\surd$\\\hline 
TT $Z_1$ IH &$\times$&$\times$&$\times$&$\times$ &$\times$&$\surd$&$\surd$&$\surd$ &$\times$&$\surd$&$\surd$&$\surd$\\\hline 
CC $Y_1$ IH &$\times$&$\times$&$\times$&$\times$ &$\times$&$\times$&$\times$&$\times$ &$\times$&$\surd$&$\times$&$\surd$\\\hline 
CC $Y_2$ NH &$\times$&$\times$&$\times$&$\times$ &$\times$&$\surd$&$\times$&$\surd$ &$\times$&$\surd$&$\times$&$\surd$\\\hline
CC $Y_2$ IH &$\times$&$\times$&$\times$&$\times$ &$\times$&$\times$&$\times$&$\times$ &$\times$&$\surd$&$\times$&$\surd$\\\hline 
CC $Z_1$ NH &$\times$&$\surd$&$\times$&$\surd$ &$\times$&$\surd$&$\times$&$\surd$ &$\times$&$\surd$&$\times$&$\surd$\\\hline 
TC 2A IH &$\times$&$\times$&$\times$&$\times$ &$\times$&$\surd$&$\times$&$\surd$ &$\times$&$\surd$&$\times$&$\surd$\\\hline 
TC 2D NH &$\times$&$\surd$&$\times$&$\surd$ &$\times$&$\surd$&$\times$&$\surd$ &$\times$&$\surd$&$\times$&$\surd$\\\hline 
TC 2D IH &$\times$&$\times$&$\times$&$\times$ &$\times$&$\times$&$\times$&$\times$ &$\times$&$\surd$&$\times$&$\surd$\\\hline 
TC 3A IH &$\times$&$\times$&$\times$&$\times$ &$\times$&$\surd$&$\times$&$\surd$ &$\times$&$\surd$&$\times$&$\surd$\\\hline 
TC 3F NH &$\times$&$\surd$&$\times$&$\surd$ &$\times$&$\surd$&$\times$&$\surd$ &$\times$&$\surd$&$\times$&$\surd$\\\hline 
TC 3F IH &$\times$&$\times$&$\times$&$\times$ &$\times$&$\times$&$\times$&$\times$ &$\times$&$\surd$&$\times$&$\surd$\\\hline 
TC 4B NH &$\times$&$\times$&$\times$&$\times$ &$\times$&$\surd$&$\times$&$\surd$ &$\times$&$\surd$&$\times$&$\surd$\\\hline 
TC 4B IH &$\surd$&$\surd$&$\times$&$\surd$ &$\surd$&$\surd$&$\times$&$\surd$ &$\surd$&$\surd$&$\times$&$\surd$\\\hline 
TC 6C IH &$\times$&$\surd$&$\surd$&$\surd$ &$\times$&$\surd$&$\surd$&$\surd$ &$\times$&$\surd$&$\surd$&$\surd$\\\hline 
\end{tabular}
\end{center}
\label{tab:survival}
\end{table}

\section*{Acknowledgments}
KW thanks the Center for Theoretical Underground Physics and Related Areas (CETUP∗
2013) in South Dakota for its hospitality and for partial support while this work was in
progress. This research was supported by the U.S. Department of Energy under Grant Nos.
DE-FG02-01ER41155, DE-FG02-13ER42024 and DE-SC0010504.

\newpage

\end{document}